# REDESIGNING TELECOMMUNICATION ENGINEERING COURSES WITH CDIO GEARED FOR POLYTECHNIC EDUCATION

*Mouhamed Abdulla, Zohreh Motamedi, and Amjed Majeed*
School of Mechanical and Electrical Engineering, Faculty of Applied Science and Technology,
Sheridan Institute of Technology, Greater Toronto Area, Ontario, Canada
Email: {mouhamed.abdulla, zohreh.motamedi, amjed.majeed} @ sheridancollege.ca

***Abstract*** *–Whether in chemical, civil, mechanical, electrical, or their related engineering subdisciplines, remaining up-to-date in the subject matter is crucial. However, due to the pace of technological evolution, information and communications technology (ICT) fields of study are impacted with much higher consequences. Meanwhile, the curricula of higher educational institutes are struggling to catch up to this reality. In order to remain competitive, engineering schools ought to offer ICT related courses that are at once modern, relevant and ultimately beneficial for the employability of their graduates. In this spirit, we were recently mandated by our engineering school to develop and design telecommunication courses with great emphasis on (i) technological modernity, and (ii) experiential learning. To accomplish these objectives, we utilized the conceive, design, implement and operate (CDIO) framework, a modern engineering education initiative of which Sheridan is a member. In this article, we chronicle the steps we took to streamline and modernize the curriculum by outlining an effective methodology for course design and development with CDIO. We then provide examples of course update and design using the proposed methodology and highlight the lessons learned from this systematic curriculum development endeavor.*

## 1. INTRODUCTION

The most populous metropolitan region in Canada, the Greater Toronto Area (GTA), is currently enjoying a large-scale technological boom. This is no surprise given that more technology-related jobs were created in the GTA than in Seattle, New York, Washington D.C., and the San Francisco Bay Area combined [1], [2]. Prominent multinational companies have either made the move to; are seriously exploring the prospect of operating their second headquarters in; or have plans to open major engineering labs in the GTA. These companies include information and communications technology (ICT) giants, such as Microsoft Corporation, Google LLC, Alphabet Inc., Intel Corporation, LG Electronics, Samsung Group, Uber Technologies, and Amazon Inc. The aim of these organizations is to remain laser-focused on pioneering innovation in artificial intelligence; machine learning; computer vision; graphical processing units; connected and autonomous transportation; and in anticipated use cases enabled by 5G communication networks.

As a consequence of this rapid technological growth, current trends suggest that Canada will be heading for a major ICT talent shortage in the next five years [3]. As such, forward-thinking engineering schools in the country are taking preventive measures. Expanding the student body and upgrading the curricula for the next wave of professionals is at the heart of this transformation. Evidently, the goal is not only to achieve the desired number of graduating engineers, but also to ensure that new graduates are adept in emerging technologies in order to meet the upcoming societal and market needs.

Meanwhile, according to industrial requirements and demands, traditional classroom learning is no longer sufficient and sustainable for skills-based employment opportunities of the near future. Expertise in prototyping, real-world engineering design, and in applied research is being sought by potential ICT employers more than ever. One particular modernization attempt in the educational system is the initiative currently underway at the School of Electrical Engineering at Sheridan Institute of Technology and Advanced Learning.

In this article, we focus on the course development activities we are undertaking in the electrical engineering program. The paper is organized as follows. In Section 2, we begin the development process by streamlining related engineering courses into a cohesive cluster; and in this case, we consider ICT related courses. Then, in Section 3, we explain the strategy and vision by which the CDIO framework is implemented for the bachelor of engineering (B.Eng.) degree program at Sheridan. Next, in Section 4, we outline a systematic methodology with CDIO guidelines for modernising, developing and designing engineering courses. Examples of course design using the proposed systematic methodology is discussed in Section 5. Finally, we provide concluding remarks in Section 6.

## 2. ORGANIZING AND STREAMLINING RELATED ENGINEERING COURSES

Before developing and designing the curricula of target courses, we find it particularly useful to streamline all telecommunication related subjects. This is necessary since many of these courses have a natural progression and build upon each other. Certainly, having a macroscopic view of ICT courses helped us to optimize with great precision the overall curriculum. Some of the advantages that we found in organizing and streamlining related courses are as follows.

(i) Clustering courses enabled us to define a unified learning objective and mission for the specialization.
(ii) It enabled us to eliminate unnecessary redundancies by minimizing the overlap of similar contents.
(iii) Interrelationship among the courses became more coherent and it gave us perspective on the topics that ought to be covered in each course.
(iv) It allowed us to reshape the course contents in a more focused and specialized manner.
(v) Optimization to the curriculum give us room to include novel engineering topics on the syllabus.
(vi) Leveraging of intellectual, technical and laboratory resources resulted in smarter cost efficient solutions.

Meanwhile, the streamlining phase was accomplished by studying the course maps for the diploma and degree programs offered by the electrical engineering department. To put things in perspective, telecommunication systems have the advantage of being structured in seven OSI layers that clearly characterize and standardize a particular technology. These layers are regrouped in one of two possible categorization, either upper or lower layer communications [4]. At Sheridan, courses with focus on the upper layers of the OSI model are generally managed by the School of Applied Computing. Other courses more aligned to the lower layers are administered by the School of Electrical Engineering. Computer engineering students are at the crossroad, which means that they take courses related to both upper and lower layer communications.

With this background in mind, once the lower layer telecommunication courses are highlighted, we envelop these specialized courses, on one side by fundamental courses, and on the other side by research courses. In fact, as shown in Fig. 1, this regrouping was made as a function of the educational learning objectives stipulated in Bloom's taxonomy [5]. Moreover, we identified a direct association to the different type of assessments, namely to:

1. <u>Diagnostic assessment</u>: related to the early stages of learning where the specific competencies are assessed before students engage in engineering projects.
2. <u>Formative assessment</u>: evaluate the particular skill sets acquired in specialized courses so as to identify areas of strength and weaknesses through gradual feedback.
3. <u>Summative assessment</u>: amalgamation of the acquired skills within the program in order to creatively conceptualize an engineering system or prototype.

## 3. IMPLEMENTATION OF THE CDIO INITIATIVE AT SHERIDAN

As highlighted earlier, within the GTA, there is great demand for ICT talent. In this region, Sheridan is one of five engineering schools offering a B.Eng. degree in mechanical and electrical engineering; the other schools are: Univ. of Toronto, York Univ., Ryerson Univ., and Ontario Tech Univ. What sets Sheridan apart is that, as a polytechnic, the objective is on skills-based learning made possible by hands-on experimental knowledge acquisition embedded within the engineering curriculum.

While this is true, the school is currently engaged in restructuring its electrical engineering program. This B.Eng. degree program and its related courses are being redesigned in such a way that it will encompass the entire life cycle of an engineering process and technology tailored precisely for the future generation of innovators and entrepreneurs. In other words, it will prepare engineering students through active learning to be involved in problem-solving skills, system design, troubleshooting, prototyping, and proof-of-concept research for real world feasibility analysis.

This educational transformation is in fact founded on the conceive, design, implement, and operate (CDIO) initiative [6], [7]. At the time of this writing, Sheridan is the only school within the GTA, and one of only five Canadian schools, alongside Polytechnique Montréal, Queen's Univ., Univ. of Manitoba, and Univ. of Calgary, that are official member schools collaborating in the CDIO initiative. Practically speaking, designing a modern B.Eng. curriculum with emphasis on experiential learning under the CDIO guidelines is not necessarily a trivial task. Yet, we attempt to approach this undertaking gradually.

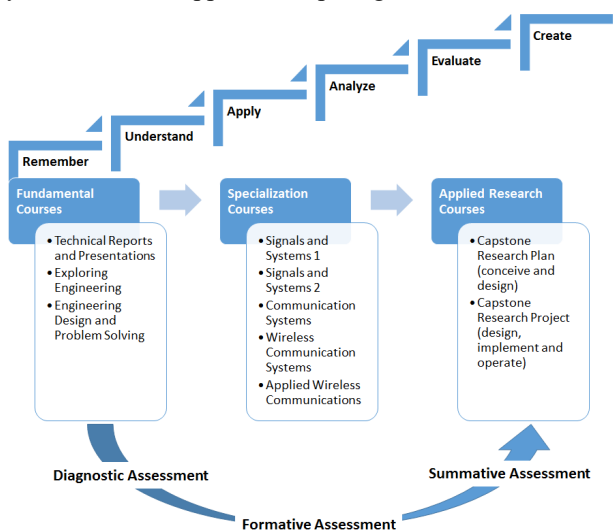

Fig. 1. Classifying ICT courses as a function of learning objectives (Bloom's taxonomy), and assessment types.

## 4. EFFECTIVE METHODOLOGY FOR COURSE DESIGN WITH CDIO

Evidently, designing modern engineering courses is a long task that will require various iterations until a desired educational outcome is achieved. Nonetheless, in this section, we propose a systematic methodology for course development and design. The technique is in part based on a combination of *backward course design*; followed by *forward course design*. In backward course design (for more on this, see [8]), we identify desired outcomes expected from students taking the course, and then plan accordingly the learning instructions, topics and modules. In forward design, we move the other way around; in other words, we progress from the course plan to the desired skill sets. In Fig. 2, we summarize, in an algorithmic style, the granularity of this approach. The steps annotated in Fig. 2, where CDIO is at the heart of this curriculum design methodology, is described as follows:

1. Skill sets: identify and list learning outcomes, and anticipated and desired skill sets acquired once the course is successfully completed.
2. Modules: determine and pinpoint pedagogical modules based on key topics desired for inclusion in the course. Then, organize the modules in such a way that the course progresses coherently for learners and educators alike.
3. Submodules: expand and elaborate every identified module with supporting technical subtopics worthy of greater consideration and explanation.
4. Lectures: determine the number of lectures needed to fully cover the submodules. This must be designed in such a way that the overall allocated lecturing time budget for the course is followed.
5. Evaluation scheme: propose a potential scheme for assessing the performance and qualification of learners. Highlight the specific learning outcomes being assessed with each evaluation item.
6. Backward design: the process of navigating from skills, to modules, to submodules, to lectures, and onto the evaluation scheme is possible due to the backward design framework.
7. CDIO guidelines: According to the learning level of the course being developed, emphasize the active learning outcomes expected from students. For instance, 1st year courses should primarily focus on I-O; 2nd and 3rd year courses on D-I-O; and 4th year courses on the C-D-I-O framework as a whole.
8. Hands-on experimentation: Prepare hands-on experimentation related to the content covered by the lecture sessions. Of course, this ought to be done as a function of the CDIO guidelines and expectations stipulated earlier.
9. Laboratory setups: Identify the laboratory apparatus needed in order to support effective CDIO-based outcomes and curriculum delivery. Such examples include the setup of hardware and software units, equipment modernization and acquisition, operating licenses, etc.

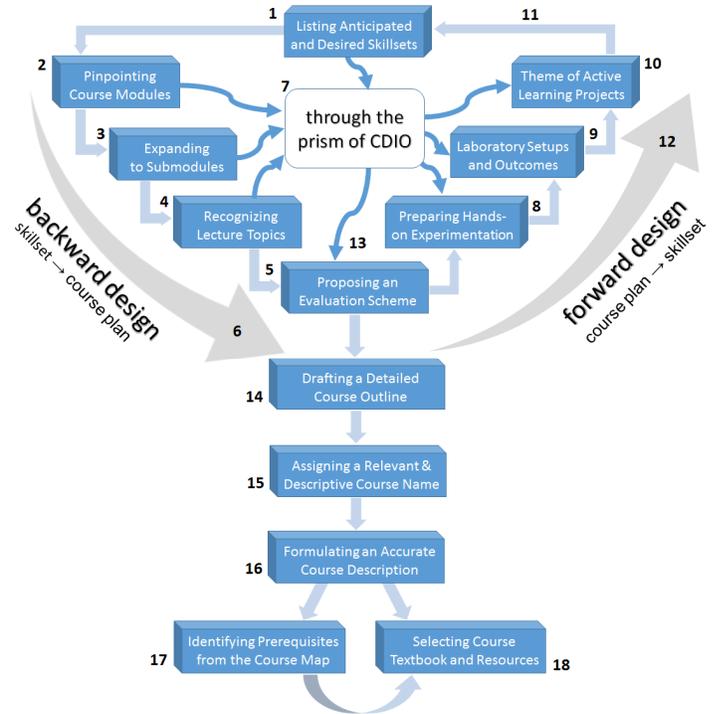

Fig. 2. Systematic course design methodology with CDIO using a combination of backward/forward design.

10. Projects: Regroup the proposed experimentations into an overarching project having a unified theme and related to a particular module. The various labs within a project can be interrelated and the learning outcomes may build on each other.
11. Updating skill sets: With the hands-on engineering component for the course established, additional skills-based applied learning items can become apparent. As a result, the desired skill sets can be updated accordingly.
12. Forward design: the process of moving from evaluations, to experiments, to laboratory setups, to projects, and onto the acquired skills is possible through the forward design process.
13. Updating evaluation scheme: with the unique proficiency acquired from hands-on CDIO-based learning and laboratory work, the evaluation scheme can at present be updated.
14. Course outline: drafting a detailed course outline will concretize the overall learning outcomes and expectations of the course. In fact, the course outline should give students a macroscopic and microscopic view of the course content and the anticipated skill sets gained once successfully completed.
15. Course name: vague and inaccurate course names should be corrected by suggesting descriptive names that are directly related to the content of the course.

16. <u>Course description</u>: with the draft of the outline ready, writing a descriptive summary for the course should be straightforward. Certainly, important skill sets, keywords and acronyms ought to be mentioned in the course description.
17. <u>Prerequisites</u>: selecting appropriate prerequisites and/or corequisites for a particular course can be done by carefully studying the course map of the program alongside the course description.
18. <u>Textbook</u>: depending on the identified prerequisites, and the streamlining of the courses described earlier, a textbook can be selected in such a way that it can be used in multiple courses. Doing so can aid learners by lessening their educational expenses.

Overall, it is interesting to notice that in this proposed methodology the process is primarily dependent on backward design, and then on forward design; not the other way around. The reason is that knowing the overreaching learning goals constitute the essentials of the course.

## 5. EXAMPLES OF COURSE DESIGN WITH EXPERIENTIAL LEARNING

In this section, we show examples of how this proposed methodology is used to develop and design engineering courses. Evidently, we modernized and embedded CDIO learning outcomes in two outdated telecommunication courses. To be exact, we developed the *Communication Systems* course, and we completely redesigned the *Wireless Communication System*s course so as to fit current advances in the field. In Tables 1 and 2, we present some of the prominent changes applied to the curricula using the course design methodology explained in Section 4 above.

We should highlight that we modified both courses in such a way that that the content, labs, projects and learning outcomes followed the integrated learning block (ILB) approach. This course structuring method (sometimes referred to as BUS-topology) essentially suggests that the allotted learning outcomes from these two courses can be transferred to an interlinked intellectual element. In other words, students will have the opportunity to apply newly acquired skill sets from different courses using the ILB approach on CDIO-based engineering projects.

## 6. CONCLUSION

In this article, we discussed the mechanism that we recently considered in order to update, reform and improve two telecommunication courses. In doing so, we first organized and streamlined all related courses so as to identify a common thread and objective. At the same time, we ensured that each course remains unique by phasing out unnecessary overlaps and redundancies. We then elaborated on the steps that we took to modernize the engineering curriculum by outlining an effective methodology for course design and development while at the same time incorporating CDIO guidelines. This was accomplished using a combination of backward and forward course design procedures. Ultimately, the objective of these changes is to facilitate the training of engineering students in practical hands-on projects through problem-solving, system design, troubleshooting and prototyping. Finally, we demonstrated examples of course design by applying the steps in the proposed methodology.

**Table 1:** Course development example applied to "Communication Systems" (ENGI-28779).

|  | before | after |
|---|---|---|
| name | Communication Systems 1 | Communication Systems |
| focus | Circuit-Level Communications (100% analog) | System-Level Communications (20% analog + 80% digital) |
| evaluation | minimum grade of 50% to pass the course | dual-pass provision in effect 60% (midterm and Final) 40% (laboratory and quizzes) |
| prerequisite | AC Circuits course | Digital Principles course |
| Laboratory | 8 labs (3.75% each) | 10 hands-on labs (3% each)<br>• kept 2 experiments on AM/FM<br>• developed 8 new digital com. labs that meet the CDIO and Integrated Learning Block (i.e. BUS-structure)<br>  ○ 5 labs emulation/prototyping<br>  ○ 3 labs simulations |
| hardware | • RF function generators<br>• RF function generators<br>• Spectrum analyzer<br>• Mixer and analog filters<br>• AM/FM radio kit | • RF function generators<br>• Spectrum analyzer<br>• Mixer and analog filters |
| software | -------------------- | Arduino IDE and MATLAB |

**Table 2:** Course design example applied to "Wireless Communication Systems" (ENGI-28649).

|  | before | after |
|---|---|---|
| name | Digital Communications | Wireless Communication Systems |
| focus | Network and some Software Layers | PHY/MAC and mostly Hardware Layers |
| evaluation | minimum grade of 50% to pass the course | dual-pass provision in effect 60% (midterm and Final) 40% (laboratory and quizzes) |
| prerequisite | Digital Principles course | Communication Systems course |
| Laboratory | 4 basic labs (7.5% each) | 4 major design projects (7.5% each)<br>• The projects and associated labs are completely redesigned to meet CDIO and Integrated Learning Block objectives.<br>• Each project is composed of 2 to 3 interrelated lab modules on wireless standards and systems. |
| hardware | • RF function generators<br>• RF function generators<br>• Spectrum analyzer<br>• Mixer and analog filters | • RF function generators<br>• Spectrum analyzer<br>• Mixer and analog filters<br>• Equipment modernization is currently underway. |
| software | MATLAB | MATLAB and Simulink |

### Acknowledgements

This pedagogical research scholarship is supported by the School of Mechanical and Electrical Engineering, Faculty of Applied Science and Technology, and Sheridan Institute of Technology.